\documentclass[conference]{IEEEtran}
\IEEEoverridecommandlockouts

\usepackage{color}
\usepackage{listings}
\definecolor{codegreen}{rgb}{0,0.67,0}
\definecolor{codegray}{rgb}{0.5,0.5,0.5}
\definecolor{codeorange}{rgb}{1,0.46,0}
\definecolor{codered}{rgb}{0.83,0,0}

\lstdefinestyle{mystyle}{ 
    commentstyle=\color{codered},
    keywordstyle=\color{codeorange},
    stringstyle=\color{codegreen},
    basicstyle=\footnotesize,
    breakatwhitespace=false,
    breaklines=true,                 
    captionpos=b,                    
    keepspaces=true,                   
    numbersep=5pt,                  
    showspaces=false,                
    showstringspaces=false,
    showtabs=false,                  
    tabsize=2
}

\usepackage[pdftex]{graphicx}

\usepackage{hyperref}

\begin{document}

\title{qgym: A Gym for Training and Benchmarking RL-Based Quantum Compilation}

\author{\IEEEauthorblockN{Stan van der Linde, Willem de Kok}
\IEEEauthorblockA{Applied Cryptography \& Quantum Algorithms\\The Netherlands Organisation for\\Applied Scientific Research (TNO)\\
The Hague, The Netherlands\\
stan.vanderlinde@tno.nl\\
willem.dekok@tno.nl}
\and
\IEEEauthorblockN{Tariq Bontekoe*}\thanks{*Work has been performed while being at TNO.}
\IEEEauthorblockA{Bernoulli Institute\\Faculty of Science and Engineering\\University of Groningen\\
Groningen, The Netherlands\\
t.h.bontekoe@rug.nl}
\and
\IEEEauthorblockN{Sebastian Feld}
\IEEEauthorblockA{Quantum \& Computer Engineering\\QuTech\\Delft University of Technology\\
Delft, The Netherlands\\
s.feld@tudelft.nl}}

\maketitle

\begin{abstract}
Compiling a quantum circuit for specific quantum hardware is a challenging task. Moreover, current quantum computers have severe hardware limitations. To make the most use of the limited resources, the compilation process should be optimized. To improve currents methods, Reinforcement Learning (RL), a technique in which an agent interacts with an environment to learn complex policies to attain a specific goal, can be used.
In this work, we present qgym, a software framework derived from the OpenAI gym, together with environments that are specifically tailored towards quantum compilation. The goal of qgym is to connect the research fields of Artificial Intelligence (AI) with quantum compilation by abstracting parts of the process that are irrelevant to either domain. It can be used to train and benchmark RL agents and algorithms in highly customizable environments.
\end{abstract}


%
\IEEEpeerreviewmaketitle

\section{Introduction}

\begin{figure}
\includegraphics[width=\linewidth]{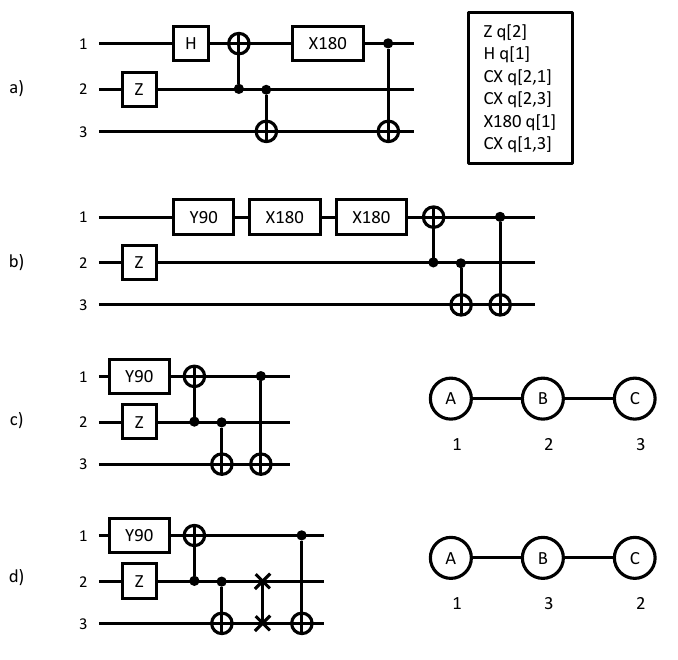}
\caption{A quantum compiler has to perform several tasks. a) An exemplary quantum algorithm given in circuit form (left-hand) and as QASM code (right-hand); b) The Hadamard gate gets decomposed into Y- and X-rotation, also the ordering of the first CNOT and the X180 gate get changed; c) The two X180 gates cancel out, the two remaining single-qubit gates are scheduled to be executed in parallel, and the logical qubits are mapped to physical qubits; d) A SWAP gate is inserted to make the last CNOT gate executable.}
\label{fig:example}
\end{figure}

In the gate-based quantum computing paradigm, qubits are controlled by applying gates to them \cite{nielsen2002quantum}.
The order and specification of these gates are often presented in a quantum circuit (see \autoref{fig:example}a together with its corresponding QASM code).
Quantum circuits offer a hardware agnostic way of thinking about gate-based quantum algorithms and there exist several frameworks that enable users to write quantum algorithms in high-level languages like C++ or Python (e.g., Qiskit \cite{Qiskit}, cirq \cite{cirq}, OpenQL \cite{khammassi2021openql}, and many more \cite{heim2020quantum}).
While the quantum circuit model is a powerful hardware agnostic tool for developing quantum algorithms \cite{weder2020quantum}, it is not at all a trivial task to execute arbitrary quantum circuits on real hardware \cite{khammassi2021openql}.
For example, a circuit might require a specific type of gate that is not implemented in the hardware, which then has to be solved by choosing a correct and efficient decomposition in available gates (see \autoref{fig:example}b, in which the Hadamard gate has been decomposed into a Y90- and an X180-rotation).
This opens up the possibility for circuit optimizations, in which commutation rules can be used to change the order of gates, followed by canceling out suitable gate pairs (see \autoref{fig:example}b, in which the first CNOT changed position with the X180 gate leading to the possibility of canceling out the two consecutive X180 gates).
Depending on the possibilities of the quantum hardware, certain gates can be scheduled to be performed at the same time (see \autoref{fig:example}c, in which the Z-gate has been moved ``under'' the Y90 gate, indicating a parallel execution).
Finally, the qubits of the circuit need to be mapped to the physical qubits of the actual hardware (see right-hand side of \autoref{fig:example}c, in which the hardware topology consists of three qubits in a row, and the logical qubits $\{1,2,3\}$ have been mapped to the physical qubits $\{A,B,C\}$, respectively).
If it turns out that the circuit cannot be executed completely in this configuration, a process called routing needs to take place, in which the logical qubits need to be moved (see \autoref{fig:example}d, in which a SWAP operation has been inserted, changing the mapping for the last CNOT and thus enabling its execution).

What the different quantum programming frameworks have in common is that they also provide means for compilation, i.e., the process of getting a quantum circuit together with a target platform as input, performing certain transformations, and outputting platform specific micro-code that is equivalent to the input circuit. The steps explained above are examples for such transformations.
The program that translates the input circuit to micro-code for the specified hardware is called a \emph{quantum compiler}.
As we are in the Noisy Intermediate-Scale Quantum (NISQ) era \cite{preskill2018quantum}, in which hardware limitations are severe, this process should be as effective as possible, making the best use of the scarce resource \cite{shi2020resource, salm2021automating}.
Additionally, compilation itself should be efficient, as the different steps taken are complex by themselves and scale poorly \cite{javadiabhari2015scaffcc}.

Reinforcement Learning (RL) is a popular concept in Artificial Intelligence (AI), in which an agent acts inside an environment by performing actions, and the environment responds to the agent with changes and giving rewards \cite{sutton2018reinforcement}.
The agents tries to maximize the reward and by doing so, it learns (potentially complex) policies that would be hard to explicitly state otherwise. RL agents are commonly trained and benchmarked in a tool called gym \cite{brockman2016openai}, in which different environments are provided to challenge them.

The goal of this paper is to present our gym, called qgym, together with some environments that are specifically tailored to quantum compilation.
Such a gym will allow and assist RL-based research on compilation-related topics such as
\begin{itemize}
    \item Qubit mapping: Learning placements of qubits that lead to a positive overall execution of an algorithm
    \item Routing: Learning strategies of moving quantum information based on a certain mapping while anticipating the future behavior of an algorithm
    \item Scheduling: Learning policies that execute algorithms while considering effects such as cross talk or memory effect
    \item Circuit optimization: learning transformation rules that might minimize the number of gates or circuit depth, or maximize fidelity
\end{itemize}
Initial work on the topics mentioned above can be seen in literature, for example \cite{cummins2021compiler,pozzi2020using,paler2023machine,fosel2021quantum}.

The main advantage of this framework is to bridge the gap between the communities of artificial intelligence and quantum computer engineering and to lower the entry barrier by hiding and abstracting parts that are irrelevant for either domain.

The contributions of this paper are as follows:
\begin{itemize}
\item We introduce the overall framework of qgym, which is based on professional software engineering principles, including thorough testing, comprehensive documentation, and diligent code commenting.
\item We provide a set of customizable environments each with states, action spaces, observation spaces, rewards and visualizations. 
These could be a starting point for RL researchers, and the environments can easily be extended.
\item We report on a proof of concept that shows that the tool indeed works as intended. This can also be used as a guideline for future users of the framework.
\end{itemize}




\section{Framework Design}

The main goal of this framework is to have a gym in which RL agents can be trained and compared.
The advantage of developing such a gym separately from a quantum compiler is twofold: 1) From the compiler point of view there is only one standardized gym API which the compiler needs to communicate with, and 2) From the point of view of the development of RL tools, the potentially irrelevant implementation details of specific quantum compilers can be  abstracted away.

\subsection{Architecture and Functionality}

The framework contains a set of environments and generic building blocks, which were used to build the environments.
As there is currently no consensus on the requirements for a ``good'' circuit and what the optimal way of compiling should look like, we put the focus of qgym on customizability.
That means that each environment has a state description, visualization tools, and a set of reward functions, all of which are customizable.
We believe that the provided set of reward functions is a good basis for a user to start with, and the framework is designed in such a way that these can easily be replaced with custom reward functions.
By doing so, we ensure that the framework provides a generic way for solving the problems encountered during quantum compilation, independently from a specific implementation.
Furthermore, qgym contains generic building blocks that were used to create the given environments. This allows to better extend the framework while keeping a uniform interface throughout all environments. The uniformity increases the viability to benchmark between environment and/or agents.

\subsection{Supported Environments}

\subsubsection{Initial mapping}

In the initial mapping problem the logical qubits of the circuit are mapped to physical qubits of the hardware such that the largest number of interactions gates in the circuit can be performed with minimal routing \cite{li2019tackling,siraichi2018qubit}.
More formally, we define a coupling graph $G_C=(V,E_C)$, where the vertices $v\in V$ represent the physical qubits of the hardware and the edges $\{v,u\}\in E_C$ represent the connections between the qubits.
Next, we define an interaction graph $G_I=(V, E_I)$, where the vertices $v\in V$ represent logical qubits of the circuit and the edges $\{v,u\}\in E_I$ represent the interactions in the circuit. 
For example, the interaction graph of the circuit given in \autoref{fig:example} would be a triangle and the connection graph are the graphs shown in  \autoref{fig:example} c) and d).
In the initial mapping problem, the objective is to find a bijection $f: V\to V$, which maps every logical qubit to a physical qubit in such a way that some cost function on the mapped edges $E_M$ is minimized, where $E_M=\big\{\{f(v),f(u)\}: \{v,u\}\in E_I\big\}$.
In some cases, a perfect mapping might exist where $E_M\subset E_C$, if this is not possible then the objective should be some cost function of $E_M \setminus E_C$, i.e., a cost function of connections that exist in the mapped interaction graph, but not in the connection graph. 
This is a generalization of the subgraph isomorphism problem, which is a well known \textsf{NP}-complete problem \cite{cook2023complexity}.

The initial mapping problem was translated to an RL environment in the following way:
The environment is initialized by a given coupling graph which remains unchanged throughout all episodes.
In each episode a new interaction graph is generated, which is observed by the agent.
Furthermore, a (initially empty) mapping is observed by the agent, which describes the bijection $f$ in the problem definition given above.
At each step, the agent can map one logical qubit to one physical qubit, until the mapping is complete.
In this way, an agent can be trained that can handle any kind of interaction graph for a predefined hardware topology.

\subsubsection{Qubit routing}
Qubit routing takes place after the initial mapping \cite{cowtan2019qubit}.
For many useful quantum algorithms a perfect initial mapping cannot be found, i.e., after the initial mapping there are still interactions in the circuit between qubits that have no connection in the connection graph. This is due to the discrepancy between the connectivity of interaction graph and coupling graph, i.e., the quantum hardware's actual topology.
Fortunately, it is still possible to execute an equivalent quantum circuit by adding swap gates to the circuit \cite{herbert2018depth}.
Graphically speaking, a swap gate exchanges two qubit lines in the circuit, circumventing the connectivity problem in the hardware topology.
However, adding swaps adds gates to the circuit and thus increases the circuit depth.
Since qubits have a short decoherence time and every gate operation introduces noise, having a short circuit depth is of huge importance for the quality of the algorithm \cite{bandic2022full}.
Finding the minimal amount of swap gates to execute the circuit is called the qubit routing problem and is known to be \textsf{NP}-hard \cite{ito2023algorithmic}.

To create an RL environment which captures the qubit routing problem, we start by constructing what we call the \emph{interaction circuit}.
The interaction circuit is a list of all two-qubit interactions ordered in the same way as the original circuit.
Similarly to the initial mapping environment, the routing environment is initialized using a given coupling graph which remains unchanged throughout all episodes.
The agent observes this topology together with (part of) the interaction circuit and optionally some additional information about the circuit.
The environment assigns a position to the agent representing the position in the interaction circuit.
This position is initialized at the start of the interaction circuit. 
At each step the agent can add a swap gate at the current position in the circuit, or move to the next position in the circuit if the coupling graph allows it.
In this manner the agent ``walks'' over the interaction circuit, adding swaps where needed.
The agent's task is finished when it reaches the end of the interaction circuit.
In this way, an agent can be trained that
can handle any kind of interaction circuit for a predefined coupling graph.

\subsubsection{Scheduling}
The scheduling problem consist of ordering in time the operations described by a (mapped, routed and decomposed) quantum circuit \cite{metodi2006scheduling}.
More precisely, each gate in the circuit should be assigned an execution cycle (the quantum hardware runs operations in cycles which are ordered time frames), such that, if the gates are executed at the assigned cycles, the output is conformable to the circuit description.
The goal is to create the shortest schedule.
Since the decoherence time of current qubit realizations is short, constructing a short schedule is of huge importance for the performance of a quantum algorithm.

The general task of scheduling is well-known, not only in quantum compilation. Scheduling as-late-as-possible (ALAP) would often give an optimal schedule.
However, when taking commutation rules of quantum gates into account, the task of scheduling becomes much more challenging.
Furthermore, there are certain hardware limitations that must be taken into account during scheduling, for example that on many systems there are certain gates that cannot be executed within the same cycle, even if they act on different qubits \cite{itoko2020scheduling}.
This form of scheduling, in which the objective is to find the shortest schedule that takes commutation rules and hardware limitations into account, is closely related to the job-shop scheduling problem, which is known to be \textsf{NP}-hard \cite{itoko2020scheduling}.

Our RL environment representing the quantum operation scheduling problem was created as follows: 
The environment is initialized with a variety of commutation rules machine properties, including the execution times of different gates and platform specific hardware limitations.
Both, the machine properties and commutation roles, do not change between episodes.
The agent starts at the end of the circuit and schedules from back to front.
At each step the agent observes the circuit and a list of gates that it currently can schedule. Furthermore, a list of dependencies is observed for each gates, and the agent can use this list to predict which gates should have priority.
The actions the agent can take in each step is either: 1) schedule a certain gate at the current cycle; or 2) move to the next cycle.
The agent is done when all gates have been scheduled.
Using this environment, an agent can be trained to schedule any circuit with pre-defined machine properties and commutation rules.

\subsection{Implementation details}

Qgym is implemented as a package in the Python programming language.
Currently, Python version 3.7 or higher are supported.
All environments use OpenAI \emph{gym} \cite{brockman2016openai} package and are compliant with the \emph{stable-baselines3} \cite{stable-baselines3} environment guidelines.
All code inside the package is tested, documented and adheres to modern coding standards. The code can be found under: \url{https://github.com/QuTech-Delft/qgym}

\section{Using qgym}

Qgym was developed as a Python package that makes the environments easy-to-use.
This section contains an example in which we show how to use an environment step-by-step, followed by a small exemplary result aiming to show possible gains of using RL methods over current methods.

\subsection{Example Usage}
\begin{figure}
\includegraphics[width=\linewidth]{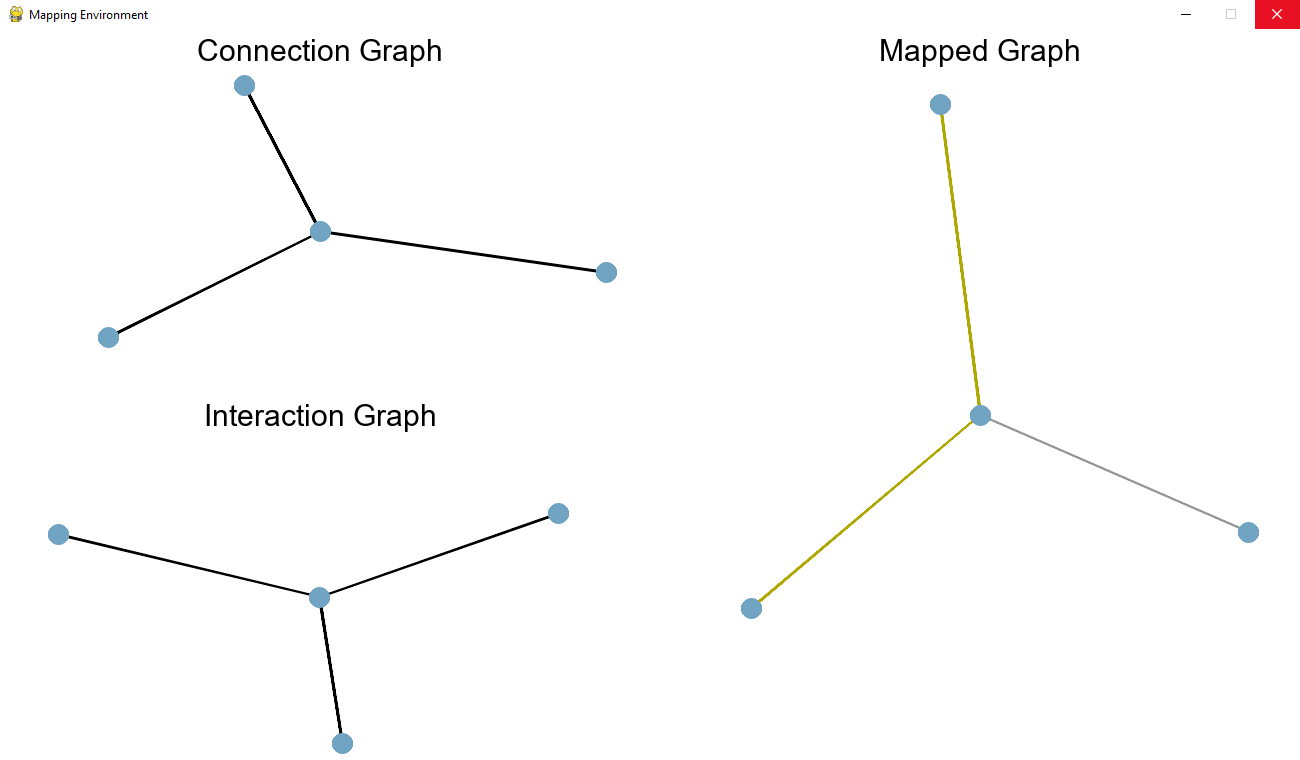}
\caption{Screenshot of the render of the initial mapping visualization.}
\label{fig:initial_mapping}
\end{figure}

As an example, we show how to create an initial mapping environment with a custom coupling graph and use this environment to train an agent.
\autoref{code1} shows how to create an initial mapping environment for a coupling graph with 4 qubits and 3 connections.
\begin{lstlisting}[language=Python, caption={Initialization of a custom InitialMapping environment}, captionpos=b, label=code1]
import networkx as nx
from qgym.envs import InitialMapping
connection_graph = nx.Graph()
connection_graph.add_edges_from(
    [(0, 1), (0, 2), (0, 3)]
)
env = InitialMapping(
    0.5, connection_graph=connection_graph
)
\end{lstlisting}
A custom agent can then be used to learn to navigate this environment.
One could develop a custom agent, or use one that is compatible with the OpenAI gym framework.
In this example an agent from the \emph{stable-baselines3} package is used.
\autoref{code2} shows how to train a vanilla  Proximal Policy Optimization (PPO) agent on the environment \cite{schulman2017proximal}.
\begin{lstlisting}[language=Python, caption={Training a PPO agent from the \emph{stable-baselines3} package on the environment initialized in \autoref{code1}},  label=code2, captionpos=b]
from stable_baselines3 import PPO
model = PPO("MultiInputPolicy", env)
model.learn(int(2e5))
\end{lstlisting}
After the agent is trained we would like to evaluate the policy learned.
\autoref{code3} shows how to make the agent use the learned policy to map a custom interaction graph.
Each step can also be rendered, see the screenshot in \autoref{fig:initial_mapping}.
\begin{lstlisting}[language=Python, caption={Let a trained PPO agent map the connection graph to itself.},  label=code3, captionpos=b]
obs = env.reset(interaction_graph=connection_graph)
for i in range(1000):
    action, states = model.predict(obs)
    obs, rewards, done, _, info = env.step(action)
    if done:
        break
\end{lstlisting}

\subsection{Experiments and Results}

\begin{figure}
\includegraphics[width=\linewidth]{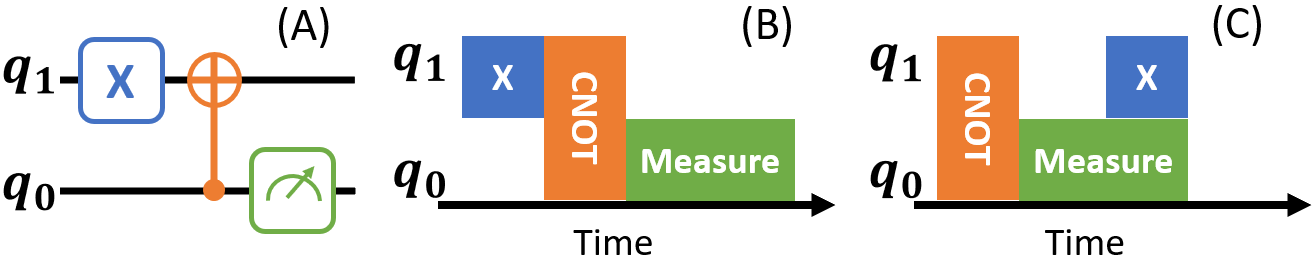}
\caption{(A) Two-qubit quantum circuit with an X gate, CNOT and measurement. (B) ALAP schedule without commutation rules. (C) Optimal schedule found by a PPO agent taking into account commutation rules.}
\label{fig:scheduling}
\end{figure}

The goal of this framework is to enable further research, however, during development we already observed that for the scheduling task even a vanilla PPO agent can learn policies which can outperform the standard priority-based ALAP approach.
We demonstrate this finding in a proof of concept.
The quantum circuit shown in \autoref{fig:scheduling}(A) will be scheduled by an ALAP method as seen is \autoref{fig:scheduling}(B).
Taking into account that the (Pauli-)X gate and CNOT in the circuit commute, the schedule shown in \autoref{fig:scheduling}(C) would be a better schedule, since it represents the same quantum circuit and is shorter in time.

The proof of concept involves showing that a PPO agent trained in our scheduling environment can produce the schedule shown in \autoref{fig:scheduling}(C).
To accomplish this, we need to define a reward function which should guide the agent to minimize the length of the schedule while avoiding illegal actions (like scheduling a operation multiple times).
We have formulated this as a reward function by giving a reward of $-5$ for an illegal action and a reward of $-1$ every time the agent increases the timestep.
All other actions receive a reward of $0$.
Thus, we chose the most basic reward function which gives a negative reward every time the agent goes to the next timestep.

\begin{figure}
\includegraphics[width=\linewidth]{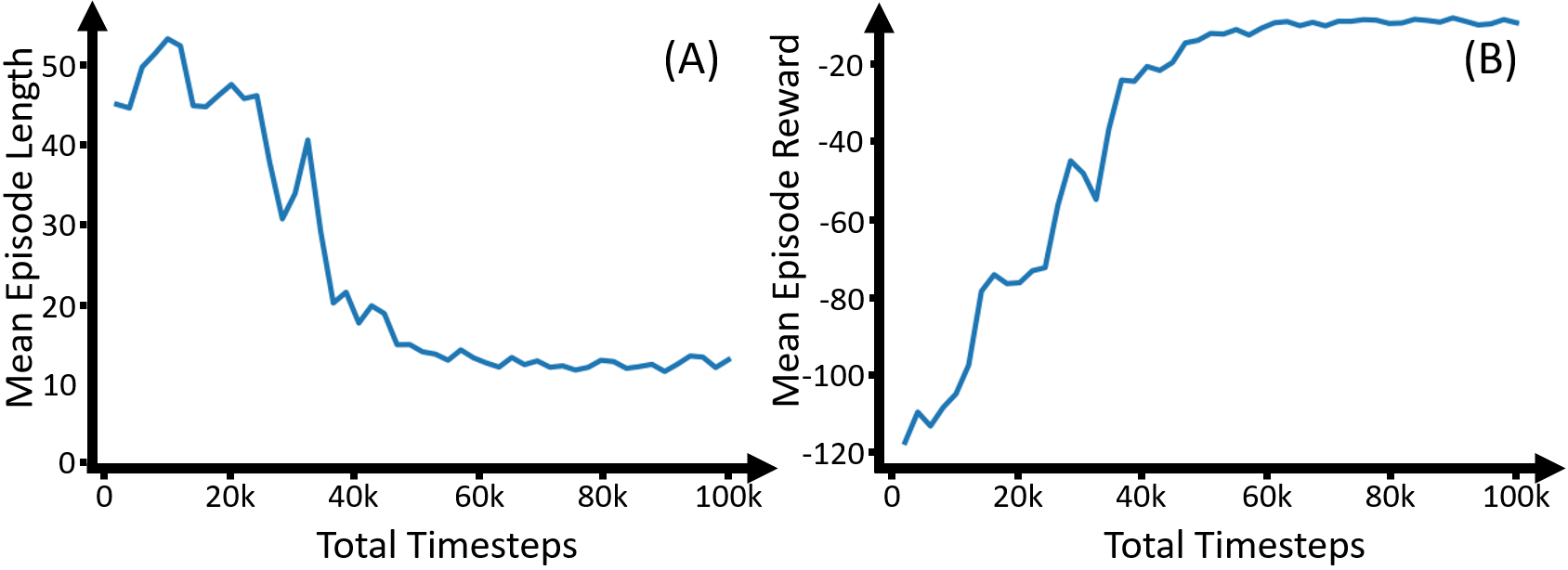}
\caption{A vanilla PPO agent has been trained in the scheduling environment. (A) The mean episode length decreases over time. (B) The mean reward per episode increases over time.}
\label{fig:learning_curve}
\end{figure}

A vanilla PPO agent from the \emph{stable-baselines3} package was used to learn a policy for a two-qubit system with commutation rules and the before mentioned reward function.
During training, a quantum circuit with at most 5 gates was randomly generated at the start of each episode.
\autoref{fig:learning_curve} shows the mean episode length and mean episode reward for the $10^5$ total steps that were taken during training.
\autoref{fig:learning_curve}(A) shows that the mean episode length decreases, while \autoref{fig:learning_curve}(B) shows an increase in reward.
This means that the number of illegal actions decreases and the length of the produced schedule decreases as well.
Hence, we gain confidence that the agent has learned a policy that minimizes the length of the schedule.
Furthermore, when the agent was given the circuit shown in \autoref{fig:scheduling}(A), it produced the schedule shown in \autoref{fig:scheduling}(C) (which is optimal).
Thus, we show that even basic non-optimized RL agents can offer improvements over a standard ALAP method.

\section{Conclusion and Outlook}

Compiling a generic quantum circuit into a hardware-specific micro-code is a hard but important task that will even continue to grow due to increasing system complexity. The concept of RL, however, can aid to train software agents in a gym to solve the compilation challenge. 

In this paper, we showcased the framework design of qgym, outlining the initial mapping, routing and scheduling environments, and how these can be used by RL researchers. Through a proof of concept, we showed that the agents can be trained to accomplish set tasks properly. 

Qgym is one of a few open source gyms for quantum compilation, therefore we invite RL researchers to train their own agents in our highly customizable gym.

As our current work focused on training agents and making the framework accessible and coherent, future work will aim at improving current environments, adding new environments, benchmarking and stress testing agents like \cite{fosel2021quantum} and also incorporating complex hardware specifics in reward functions, like the fidelity of qubit connections.

\section*{Acknowledgment}

The authors are grateful for funding by QuTech and TNO.

\bibliographystyle{IEEEtran}
\bibliography{bibliography}

\end{document}